\documentclass[aps,prb,reprint,superscriptaddress,showpacs,floatfix,citeautoscript]{revtex4-1}
\usepackage{graphicx}
\usepackage{bm}
\usepackage{dcolumn}
\usepackage{amsfonts}
\usepackage{amsmath}
\usepackage{amssymb}
\usepackage{hyperref}
\usepackage[usenames]{color}

\hypersetup{pdfborder=0 0 0,colorlinks=true, citecolor=blue,linkcolor= blue}
\begin{document}

\title{Ni(111)$|$Graphene$|$h-BN Junctions as Ideal Spin Injectors}
\author{V. M. Karpan}
\affiliation{Faculty of Science and Technology and MESA$^+$ Institute
for Nanotechnology, University of Twente, P.O. Box 217, 7500 AE Enschede, 
The Netherlands}
\author{P. A. Khomyakov}
\thanks{Present address: IBM Research - Zurich.}
\affiliation{Faculty of Science and Technology and MESA$^+$ Institute
for Nanotechnology, University of Twente, P.O. Box 217, 7500 AE Enschede, 
The Netherlands}
\author{G. Giovannetti}
\thanks{Present address:  Univ Roma La Sapienza, Dipartimento Fis, ISC CNR, I-00185 Rome, Italy .}
\affiliation{Faculty of Science and Technology and MESA$^+$ Institute
for Nanotechnology, University of Twente, P.O. Box 217, 7500 AE Enschede, 
The Netherlands} 
\affiliation{Instituut-Lorentz for Theoretical Physics, 
Universiteit Leiden, P. O. Box 9506, 2300 RA Leiden, The Netherlands}

\author{A. A. Starikov}
\affiliation{Faculty of Science and Technology and MESA$^+$ Institute
for Nanotechnology, University of Twente, P.O. Box 217, 7500 AE Enschede, 
The Netherlands}
\author{P. J. Kelly}
\affiliation{Faculty of Science and Technology and MESA$^+$ Institute
for Nanotechnology, University of Twente, P.O. Box 217, 7500 AE Enschede, 
The Netherlands}

\date{\today ~version v1c}

\begin{abstract}

Deposition of graphene on top of hexagonal boron nitride (h-BN) was very recently demonstrated while graphene is now routinely grown on Ni. Because the in-plane lattice constants of graphite, h-BN, graphite-like BC$_2$N and of the close-packed surfaces of Co, Ni and Cu match almost perfectly, it should be possible to prepare ideal interfaces between these materials which are respectively, a semimetal, insulator, semiconductor, ferromagnetic and nonmagnetic metals. Using parameter-free energy minimization and electronic transport calculations, we show how h-BN can be combined with the perfect spin filtering property of Ni$|$graphite and Co$|$graphite interfaces to make perfect tunnel junctions or ideal spin injectors (SI) with any desired resistance-area product.

\end{abstract}


%
%
%
%
%
%
%


\maketitle

{\em \color{red} Introduction.---}Progress in increasing the storage capacity of magnetic hard disk drives \cite{Wood:jmmm09} depends on finding materials with large magnetoresistance (MR) ratios and suitable resistance-area ($RA$) products for use as read-head sensors. \cite{Nagasaka:jmmm09} Current read-head technology is based upon magnetic tunnel junctions (MTJ) with polycrystalline MgO barriers and tunneling magnetoresistance (TMR) ratios MR $=(R_{\rm AP} - R_{\rm P})/R_{\rm P} \equiv (G_{\rm P} - G_{\rm AP})/G_{\rm AP}$ of around 100\%. \cite{Wood:jmmm09}  The subscripts P and AP refer to magnetizations of adjacent magnetic layers being aligned parallel and antiparallel, respectively, in this so-called {\it optimistic} definition of magnetoresistance. MTJs must satisfy a large number of constraints relating to the assembly and processing of read-heads so CoFeB electrodes are used even though larger MR ratios could be realized with other choices. \cite{Yuasa:jpd07} To maintain acceptable data-transfer rates and signal-to-noise ratios \cite{Nagasaka:jmmm09,*Gao:jmmm09} when reducing the read-head dimensions to read smaller magnetic bits, it is essential to have a small $RA$ product; to achieve bit densities of 1 Tera-bit/in$^2$, it is estimated that $RA$ should be of order $\sim 0.1 \, \Omega \mu {\rm m}^2$.\cite{Nagasaka:jmmm09} This can be achieved by making the MgO barrier thinner, but only at the cost of a reduced MR ratio. \cite{Yuasa:jpd07} Currently used tunnel junctions are so thin ($t_{\rm MgO} \sim 1.0 $ nm or only 4-5 atomic layers thick) that further reduction will introduce pinholes. If the area $A$ is to be made even smaller, it will become necessary to find systems with equally high or higher MR ratios and lower $RA$ products. Another common scenario is to use current-perpendicular-to-the-plane (CPP) metallic GMR (giant MR) sensors with MR ratios of order 100\%. 

We recently showed\cite{Karpan:prl07,*Karpan:prb08} that a very few atomic layers of graphite sandwiched between close-packed Ni or Co electrodes should have an infinite magnetoresistance. The reason is that graphene and graphite have the same in-plane lattice constant as close-packed surfaces of Ni and Co so they share a common two dimensional reciprocal space. In this reciprocal space, the states graphite has at or close to the Fermi energy are located around the K point. Ni and Co have no majority spin states at the Fermi level at the K point so majority spin states cannot enter graphite from Ni or Co without a (large) change of transverse crystal momentum. As a consequence, the majority spin conductance is attenuated exponentially when the number of graphene sheets (Gr) is increased. Ni and Co minority spin states at the Fermi level occupy all of the reciprocal space so they can enter graphite and once they couple to the Bloch states in graphite, they are not attenuated when its thickness is increased. A FM$|$Gr$_n|$FM junction has a very low $RA$ product ($\sim 0.1 \, \Omega \mu {\rm m}^2$) which depends very weakly on the number $n$ of graphene sheets. This intriguing behaviour depends upon the happy coincidence that the K points of all three materials coincide which in turn results from a near perfect matching of the in-plane lattice parameters of graphite and close packed surfaces of Ni and Co. Though this prediction of perfect spin filtering has yet to be experimentally confirmed, the physical principles upon which it is based are very well established and robust. Detailed calculations show a remarkable insensitivity to interface roughness, disorder and lattice mismatch.\cite{Karpan:prl07,*Karpan:prb08} Because it is based upon (111) oriented metal electrodes, an FM$|$Gr$_n|$FM spin filter is compatible with the industry-standard (111)-orientation ``pinned layer'' consisting of a synthetic ferrimagnetic structure and antiferromagnetic exchange biasing layer used in all read-heads. \cite{Yuasa:jpd07}

\begin{figure}[t!]
\includegraphics[scale=0.65]{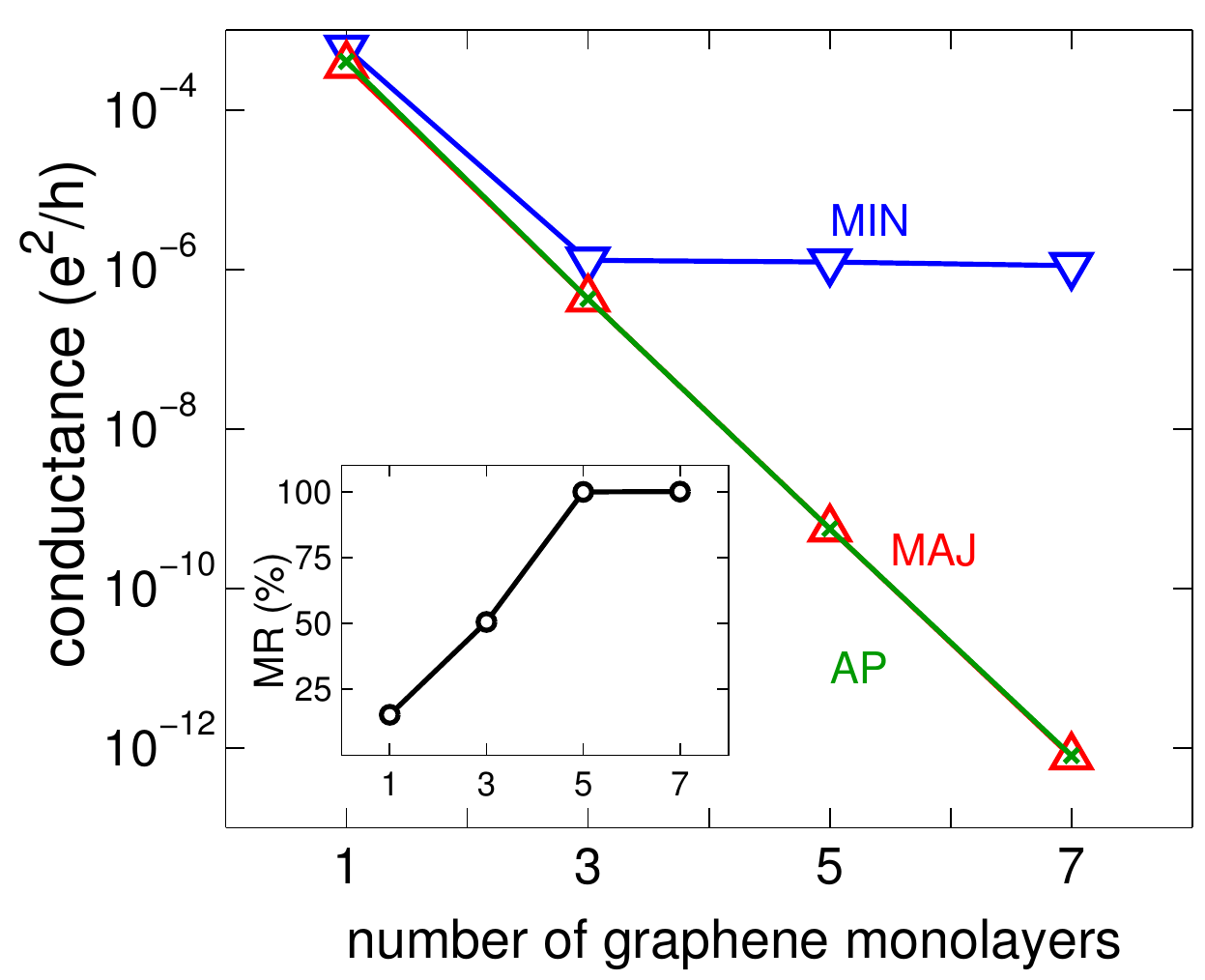}
\caption{Conductances
$G_{\rm P}^{\rm min}$ ($\triangledown$),
$G_{\rm P}^{\rm maj}$ ($\vartriangle$), and
$G_{\rm AP}^{\protect\sigma}$ (${\mathbf{\times}}$)
of a Ni$|$BN$|$Gr$_n|$Ni junction as a function of the number of sheets of graphene $n$. Results for 2 sheets of BN are shown. 
Inset: magnetoresistance as a function of $n$ for ideal junctions. Here, the {\it pessimistic} definition, MR 
$=(R_{\rm AP} - R_{\rm P})/R_{\rm AP} \equiv (G_{\rm P} - G_{\rm AP})/G_{\rm P}$ which does not diverge when $G_{\rm AP}$ is zero, is used.}
\label{ref:Fig1}
\end{figure}

{\em \color{red} Increasing $RA$.---}For other applications it is desirable to have a large $RA$ product. For example, high density magnetoresistive random access memories (MRAMs)\cite{Kryder:ieeem09} require MR ratios in excess of 150\% at room temperature, $RA$ products in the range $50 \, \Omega \mu {\rm m}^2$ to $10 \,{\rm k} \Omega \mu {\rm m}^2$ depending on the areal density, and lower current densities for switching via spin-transfer torque.\cite{Yuasa:jpd07} Spin electronics or ``spintronics'' aims to introduce into conventional semiconductor-based electronics the additional spin degree of freedom used to such good effect in metal-based ``magnetoelectronics''. Attempts to inject spins directly into semiconductors encounter a so-called ``conductivity mismatch'' problem: the difference in spin-up and spin-down resistivity in conventional ferromagnetic metals is negligible compared to the very much larger spin-independent resistivity of semiconductors.\cite{Schmidt:prb00} This problem can be resolved by injecting spins through a spin-dependent tunnel junction \cite{Fert:prb01} or Schottky barrier \cite{Hanbicki:apl02} but up till now the spin polarizations achieved at room temperature are far from complete. Here we show how the $RA$ product of an FM$|$Gr$_n|$FM junction with FM = Ni or Co, can be made arbitrarily large without reducing the polarization, by inserting $m$ sheets of h-BN to make an FM$|$BN$_m|$Gr$_n|$FM(111) MTJ. \cite{Karpan08,[][{ have considered the cases $(m,n)=$ (0,1) and (1,0) with monolayers of graphene or h-BN sandwiched between fcc Fe, Co or Ni or hcp Co electrodes.}]Yazyev:prb09,[][{ have considered the case (2,0) with 2 layers of h-BN sandwiched between Ni electrodes.}]Hu:jpcc11} 
Hexagonal BN is a large band gap semiconductor with an indirect gap of $6$~eV.\cite{Arnaud:prl06} More importantly, it has the same honeycomb structure as graphene, almost the same lattice parameter, and can be prepared in monolayer form by micromechanical cleavage.\cite{Novoselov:pnas05} Recent success in preparing graphene on top of h-BN has led to the observation of mobilities comparable to those observed in freely suspended graphene and has opened the way to a host of new transport studies.\cite{Dean:natn10,*Xue:natm11,*Decker:nanol11} In Fig.\ref{ref:Fig1} we show that inserting two layers of h-BN between the Ni electrode and graphene sheets increases the $RA$ product by more than three yields of magnitude \cite{Karpan:prl07,*Karpan:prb08} without any deterioration in the polarization. In both cases, with and without h-BN, an ideal, essentially 100\% MR ratio (pessimistic definition, MR $=(R_{\rm AP} - R_{\rm P})/R_{\rm AP} \equiv (G_{\rm P} - G_{\rm AP})/G_{\rm P}$) is achieved with five layers of graphene. \cite{Karpan:prl07,*Karpan:prb08} By adjusting the number of layers of Gr and h-BN, the $RA$ product can be varied essentially arbitrarily. Relaxed structures were determined {\em ab-initio} by energy minimization \cite{Giovannetti:prb07} using a plane-wave basis set and the Projector Augmented Wave (PAW) formalism\cite{Blochl:prb94b} as implemented in the VASP code.\cite{Kresse:prb93,*Kresse:prb96,*Kresse:prb99}  Conductances were calculated from first-principles using the tight-binding muffin-tin-orbital wavefunction matching method \cite{Xia:prb06} used in our previous studies. \cite{Karpan:prl07,*Karpan:prb08}

\begin{figure}[t!]
\includegraphics[scale=0.65]{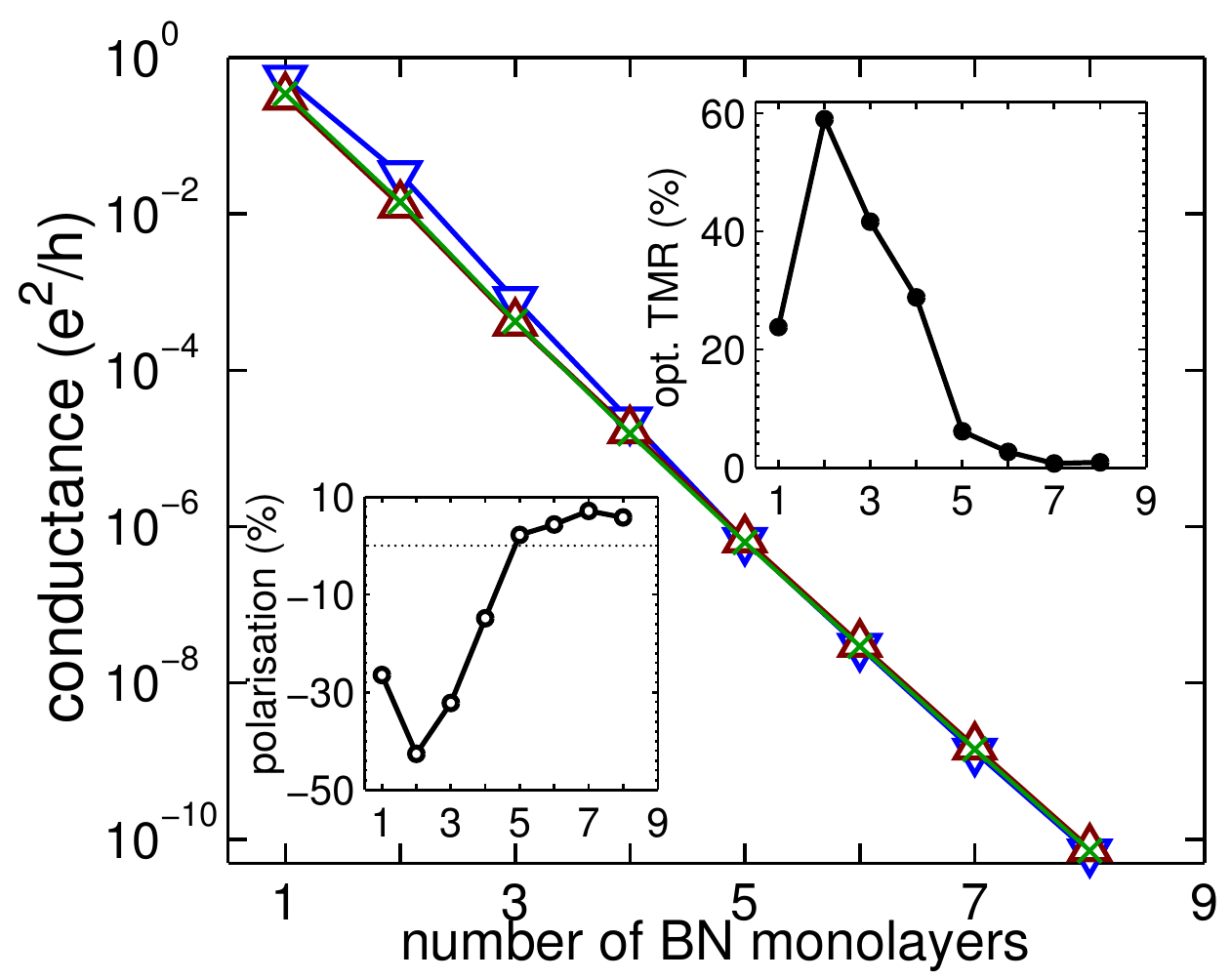}
\caption{Conductances
$G_{\rm P}^{\rm min}$ ($\triangledown$),
$G_{\rm P}^{\rm maj}$ ($\vartriangle$), and
$G_{\rm AP}^{\protect\sigma}$ (${\mathbf{\times}}$) of a Ni$|$BN$_n|$Ni
junction as a function of the number of BN layers $n$ for ideal junctions.
Insets: {\it{optimistic}} magnetoresistance as a function of $n$ for ideal junction (on the right) and polarization of the parallel conductance $P=(G_{\rm P}^{\rm maj}-G_{\rm P}^{\rm min})/(G_{\rm P}^{\rm maj}+G_{\rm P}^{\rm min})$.}
\label{ref:Fig2}
\end{figure}

The large magnetoresistance of Fe$|$MgO$|$Fe MTJs \cite{Butler:prb01a,Mathon:prb01,Parkin:natm04,Yuasa:natm04,Yuasa:jpd07} is attributed to the crystallinity of the MgO tunnel barrier. Since h-BN is crystalline and its in-plane lattice constant matches those of (111) Ni and Co to better than a percent, we investigated the magnetoresistance of FM$|$BN$_m|$FM(111) MTJs with $m$ sheets of h-BN.\cite{Karpan08,Yazyev:prb09,Hu:jpcc11} The results are shown in Fig.~\ref{ref:Fig2} for Ni electrodes. It can be seen that in the wide barrier limit the magnetoresistance vanishes. The small MR found for thin barriers can be traced to the existence of a surface state in the minority channel on Ni(111). As the barrier width increases, the contribution from this surface state is quenched. MgO is a cubic material with a conduction band minimum at the $\Gamma$-point that is much lower in energy than at other high symmetry points. States at the Fermi energy of the metal electrode which match the $s$-like symmetry of this conduction band minimum are attenuated much more slowly in MgO than states with other symmetries. In the case of Fe, there is a state with this orbital character at the Fermi energy for majority spin but not for minority spin. By contrast, the bottom of the conduction band (top of the valence band) of h-BN at the K, $\Gamma$, M, H and L (respectively, K, $\Gamma$, M, H, A and L) high symmetry points in reciprocal space have very similar energies \cite{Arnaud:prl06} so that there is no preferential tunnelling of states with a particular orbital character which might translate as in the Fe$|$MgO(001) case into preference for a particular spin channel. Perfect lattice matching alone is not enough to obtain a large magnetoresistance and h-BN must be used in combination with graphite to simultaneously obtain a high MR ratio and large $RA$ product.

{\em \color{red} Ideal Spin Injector.---}The perfect spin filtering properties of graphite on close-packed surfaces of Ni or Co means that this hybrid system, which we denote FM(111)$|$Gr$_n$, behaves as a half-metallic material and can be used to inject a 100\% spin-polarized current into non-magnetic materials. As an example, we consider spin injection into metallic aluminium where till now the most successful means of injecting spin has been by using an aluminium oxide tunnel barrier.\cite{Jedema:nat02}

\begin{figure}[tbp]
\includegraphics[scale=0.40,angle=0]{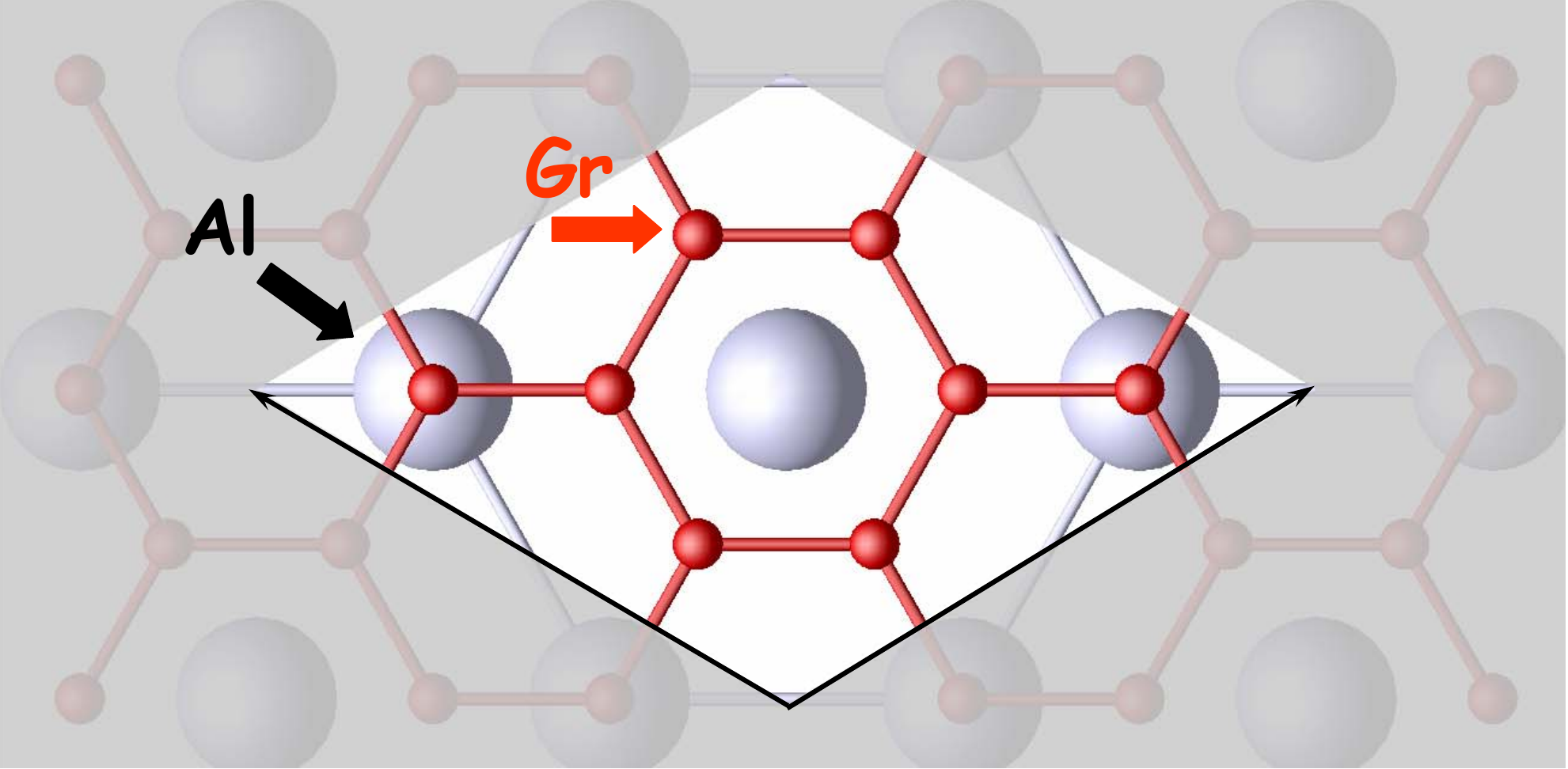}
\caption{The most stable configuration of graphene on (111) surfaces of the fcc non-magnetic metals Al, Ag, Au, Pd and Pt with two carbon atoms on top of metal atoms and the remaining six  on bridge sites. The interface unit cell contains 8 carbon and 3 metal atoms.}
\label{ref:Fig3}
\end{figure}

\begin{figure}[tbp]
\includegraphics[scale=0.6]{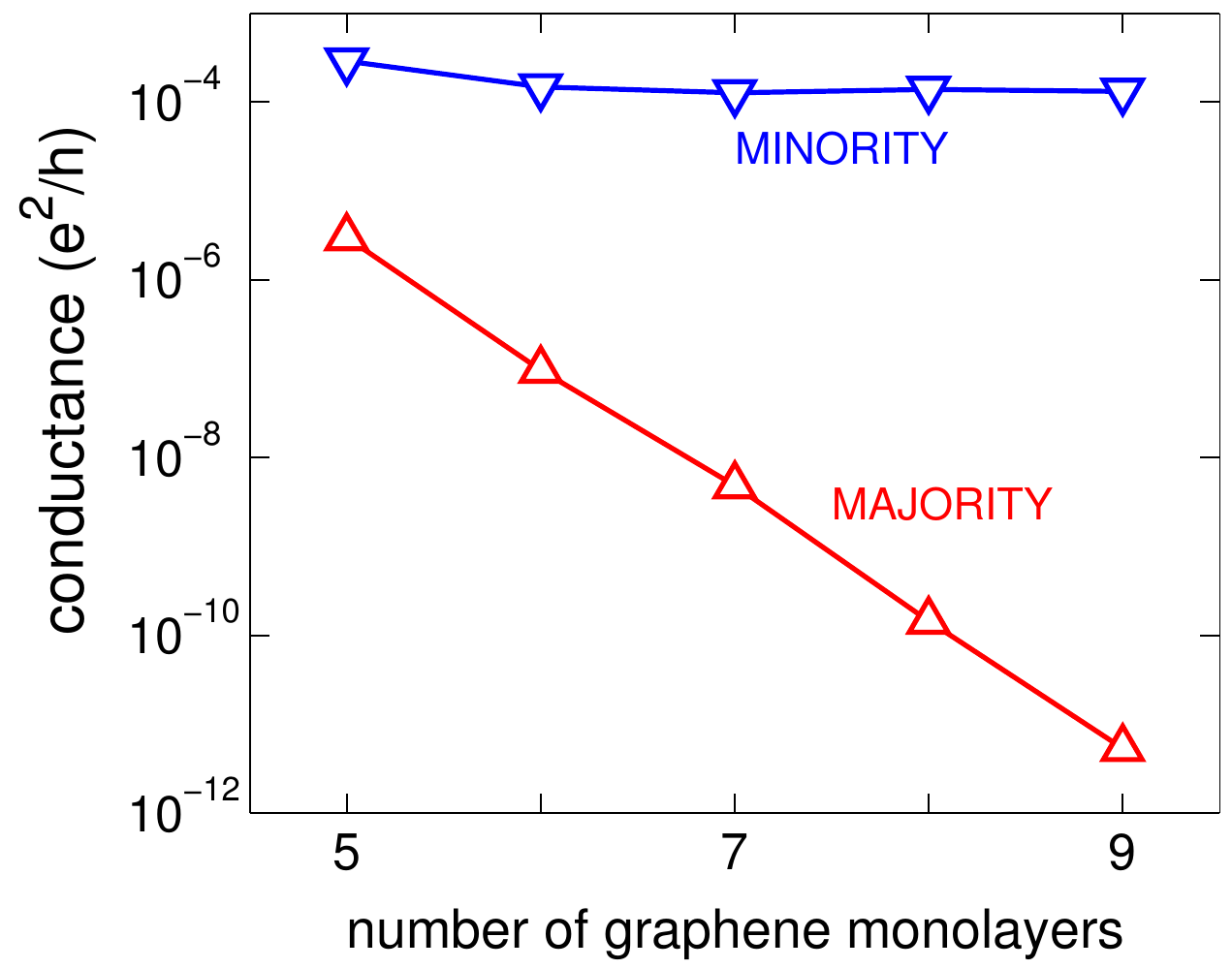}
\caption{Conductances $G^{\rm min}$ ($\triangledown$), $G^{\rm maj}$
($\vartriangle$) of a Ni$|$Gr$_n|$Al junction as a function of the
number of graphene layers $n$ for ideal junctions. } 
\label{ref:Fig4}
\end{figure}

The lattice constants of the face-centred cubic non-magnetic metals (NM) Al, Ag, Au, Pd and Pt are such that a $2 \times2 $ unit cell of graphene containing 8 carbon atoms matches a $\sqrt{3} \times \sqrt{3}$ surface unit cell of the (111) non-magnetic metal almost perfectly.\cite{Giovannetti:prl08,*Khomyakov:prb09} The most stable configuration of a graphene$|$NM(111) interface is determined without introducing free parameters by minimizing the density functional theory (DFT) total energy within the local density approximation. The lowest energy  symmetric configuration we found is illustrated in Fig.~\ref{ref:Fig3}. It has two carbon atoms above NM atoms while the remaining six occupy ``bridge'' sites between metal atoms. The equilibrium interlayer distance at the interface is calculated to be $d=3.41 \AA$ for aluminium. For the other non-magnetic metals Ag, Au, Pd, and Pt, equilibrium geometries and binding energies can be found elsewhere. \cite{Giovannetti:prl08,*Khomyakov:prb09} Once an interface geometry has been determined, the conductance can be calculated. The results are shown in Fig.~\ref{ref:Fig4} for Ni$|$Gr$_n|$Al(111) as a function of $n$. This figure demonstrates the saturation of the minority spin injection and rapid exponential attenuation of the majority spin injection resulting in 100\% polarization of the injected carriers.

FM(111)$|$Gr$_n$ could also be used to inject spins into a doped semiconductor such as Si or GaAs though in practice it might be desirable to include some layers of h-BN to match the impedance of the semiconductor. Should the 6~eV bandgap of h-BN be too large, BC$_2$N with the same layered structure as graphite and h-BN and a bandgap of 2~eV \cite{Watanabe:prl96,Chen:prl99} slightly larger than those of Si and GaAs is an alternative.

{\em \color{red} Discussion.---}The spin filtering discussed in the preceding is based upon the materials considered having lattice constants which are almost perfectly matched so that they share a common two dimensional reciprocal space where only states of one spin in (111)-oriented Ni (or Co) are compatible with the states at the Fermi energy in graphene or graphite.\cite{Karpan:prl07,*Karpan:prb08} Since the calculations we have presented would have been impossible without this lattice matching, this poses the question of how sensitive our results will be to any perturbation which breaks perfect translational symmetry. In Ref.~\onlinecite{Karpan:prl07,*Karpan:prb08}, we explicitly studied the effect of disorder and lattice-mismatch at one of the interfaces on the spin filtering and found it to be small, much smaller than in the case of Fe(001)$|$semiconductor interface disorder \cite{Zwierzycki:prb03} or interface roughness for MTJs.\cite{Xu:prb06a} We argued that the robustness of the FM$|$Gr$_n|$FM(111) spin filtering to these types of disorder was related to the large region about the K-point in reciprocal space where there are no Ni (or Co) majority-spin states. The same arguments should hold when h-BN is included in a FM$|$Gr$_n|$BN$_m|$FM(111) MTJ. 

For FM(111)$|$Gr$_n$ or FM$|$Gr$_n|$BN$_m$ to be useful as spin injectors (SI), the non-magnetic material (NMM) into which we wish to inject spins should ideally not have to be lattice matched to the injector. We expect the nature of the SI$|$NMM interface or of NMM itself (whether or not it is lattice-matched or even crystalline) to be unimportant as long as (i) the FM(111)$|$Gr$_n$ interface is sufficiently defect-free that the majority spin interface resistance is large compared to $\rho_c \ell_c$, the $c$-axis resistivity of graphite times the mean free path in graphite and (ii) $n$ is large, $i.e.$ the graphite layer is so thick that the FM(111)$|$Gr$_n$ and SI$|$NMM interfaces are essentially uncoupled so there is no tunnelling of majority spin electrons through graphite into NMM. In this case, it should be possible to consider electronic transport in a two step process. In the first step, spins are injected from FM(111) into graphite creating a spin-accumulation there. In the second step, the non-equilibrium spins are injected into the NMM and the conservation of transverse momentum is unimportant. Though many studies indicate that it is possible to prepare essentially perfect Ni$|$graphite interfaces, quantitative studies need to be made of the maximum majority spin interface resistance achievable for a given interface defect density. If they can be made sufficiently defect-free, then it should be possible to establish a non-equilibrium spin population of the conduction bands of graphite (``spin-accumulation'') which will have a long lifetime because of the low atomic number of carbon and its very weak intrinsic spin-orbit interaction. 

{\em \color{red} Summary.---}Lattice matched materials figure prominently in the discovery of new physical effects at interfaces. Because the in-plane lattice constants of semimetallic graphite, insulating hexagonal boron nitride (h-BN), semiconducting BC$_2$N and of the close-packed surfaces of ferromagnetic Co and Ni and non-metallic Cu match almost perfectly, these materials form an interesting system in which to study electronic transport. We showed by explicit calculation how h-BN can be used to increase the $RA$ product of FM$|$Gr$_n|$FM(111) spin filters without decreasing the magnetoresistance and how FM$|$Gr$_n|$ could be used to inject a single spin species into Al (a superconductor at low temperatures) as an example of a non-magnetic material. 


\begin{acknowledgments}
This work was supported by ``NanoNed'', a
nanotechnology programme of the Dutch Ministry of Economic Affairs. It
is part of the research programs of ``Chemische Wetenschappen'' (CW) and
``Stichting voor Fundamenteel Onderzoek der Materie'' (FOM) and the use of
supercomputer facilities was sponsored by the ``Stichting Nationale Computer
Faciliteiten'' (NCF), all financially supported by the ``Nederlandse
Organisatie voor Wetenschappelijk Onderzoek'' (NWO).
\end{acknowledgments}


\begin{thebibliography}{35}%
\makeatletter
\providecommand \@ifxundefined [1]{%
 \@ifx{#1\undefined}
}%
\providecommand \@ifnum [1]{%
 \ifnum #1\expandafter \@firstoftwo
 \else \expandafter \@secondoftwo
 \fi
}%
\providecommand \@ifx [1]{%
 \ifx #1\expandafter \@firstoftwo
 \else \expandafter \@secondoftwo
 \fi
}%
\providecommand \natexlab [1]{#1}%
\providecommand \enquote  [1]{``#1''}%
\providecommand \bibnamefont  [1]{#1}%
\providecommand \bibfnamefont [1]{#1}%
\providecommand \citenamefont [1]{#1}%
\providecommand \href@noop [0]{\@secondoftwo}%
\providecommand \href [0]{\begingroup \@sanitize@url \@href}%
\providecommand \@href[1]{\@@startlink{#1}\@@href}%
\providecommand \@@href[1]{\endgroup#1\@@endlink}%
\providecommand \@sanitize@url [0]{\catcode `\\12\catcode `\$12\catcode
  `\&12\catcode `\#12\catcode `\^12\catcode `\_12\catcode `\%12\relax}%
\providecommand \@@startlink[1]{}%
\providecommand \@@endlink[0]{}%
\providecommand \url  [0]{\begingroup\@sanitize@url \@url }%
\providecommand \@url [1]{\endgroup\@href {#1}{\urlprefix }}%
\providecommand \urlprefix  [0]{URL }%
\providecommand \Eprint [0]{\href }%
\providecommand \doibase [0]{http://dx.doi.org/}%
\providecommand \selectlanguage [0]{\@gobble}%
\providecommand \bibinfo  [0]{\@secondoftwo}%
\providecommand \bibfield  [0]{\@secondoftwo}%
\providecommand \translation [1]{[#1]}%
\providecommand \BibitemOpen [0]{}%
\providecommand \bibitemStop [0]{}%
\providecommand \bibitemNoStop [0]{.\EOS\space}%
\providecommand \EOS [0]{\spacefactor3000\relax}%
\providecommand \BibitemShut  [1]{\csname bibitem#1\endcsname}%
\let\auto@bib@innerbib\@empty
\bibitem [{\citenamefont {Wood}(2009)}]{Wood:jmmm09}%
  \BibitemOpen
  \bibfield  {author} {\bibinfo {author} {\bibfnamefont {R.}~\bibnamefont
  {Wood}},\ }\href {\doibase 10.1016/j.jmmm.2008.07.027} {\bibfield  {journal}
  {\bibinfo  {journal} {J. Magn. \& Magn. Mater.}\ }\textbf {\bibinfo {volume}
  {321}},\ \bibinfo {pages} {555} (\bibinfo {year} {2009})}\BibitemShut
  {NoStop}%
\bibitem [{\citenamefont {Nagasaka}(2009)}]{Nagasaka:jmmm09}%
  \BibitemOpen
  \bibfield  {author} {\bibinfo {author} {\bibfnamefont {K.}~\bibnamefont
  {Nagasaka}},\ }\href {\doibase 10.1016/j.jmmm.2008.05.040} {\bibfield
  {journal} {\bibinfo  {journal} {J. Magn. \& Magn. Mater.}\ }\textbf {\bibinfo
  {volume} {321}},\ \bibinfo {pages} {508} (\bibinfo {year}
  {2009})}\BibitemShut {NoStop}%
\bibitem [{\citenamefont {Yuasa}\ and\ \citenamefont
  {Djayaprawira}(2007)}]{Yuasa:jpd07}%
  \BibitemOpen
  \bibfield  {author} {\bibinfo {author} {\bibfnamefont {S.}~\bibnamefont
  {Yuasa}}\ and\ \bibinfo {author} {\bibfnamefont {D.~D.}\ \bibnamefont
  {Djayaprawira}},\ }\href {\doibase 10.1088/0022-3727/40/21/R01} {\bibfield
  {journal} {\bibinfo  {journal} {J. Phys. D: Appl. Phys.}\ }\textbf {\bibinfo
  {volume} {40}},\ \bibinfo {pages} {R337} (\bibinfo {year}
  {2007})}\BibitemShut {NoStop}%
\bibitem [{\citenamefont {Gao}\ \emph {et~al.}(2009)\citenamefont {Gao},
  \citenamefont {Heinonen},\ and\ \citenamefont {Chen}}]{Gao:jmmm09}%
  \BibitemOpen
  \bibfield  {author} {\bibinfo {author} {\bibfnamefont {K.~Z.}\ \bibnamefont
  {Gao}}, \bibinfo {author} {\bibfnamefont {O.}~\bibnamefont {Heinonen}}, \
  and\ \bibinfo {author} {\bibfnamefont {Y.}~\bibnamefont {Chen}},\ }\href
  {\doibase 10.1016/j.jmmm.2008.05.025} {\bibfield  {journal} {\bibinfo
  {journal} {J. Magn. \& Magn. Mater.}\ }\textbf {\bibinfo {volume} {321}},\
  \bibinfo {pages} {495} (\bibinfo {year} {2009})}\BibitemShut {NoStop}%
\bibitem [{\citenamefont {Karpan}\ \emph {et~al.}(2007)\citenamefont {Karpan},
  \citenamefont {Giovannetti}, \citenamefont {Khomyakov}, \citenamefont
  {Talanana}, \citenamefont {Starikov}, \citenamefont {Zwierzycki},
  \citenamefont {van~den Brink}, \citenamefont {Brocks},\ and\ \citenamefont
  {Kelly}}]{Karpan:prl07}%
  \BibitemOpen
  \bibfield  {author} {\bibinfo {author} {\bibfnamefont {V.~M.}\ \bibnamefont
  {Karpan}}, \bibinfo {author} {\bibfnamefont {G.}~\bibnamefont {Giovannetti}},
  \bibinfo {author} {\bibfnamefont {P.~A.}\ \bibnamefont {Khomyakov}}, \bibinfo
  {author} {\bibfnamefont {M.}~\bibnamefont {Talanana}}, \bibinfo {author}
  {\bibfnamefont {A.~A.}\ \bibnamefont {Starikov}}, \bibinfo {author}
  {\bibfnamefont {M.}~\bibnamefont {Zwierzycki}}, \bibinfo {author}
  {\bibfnamefont {J.}~\bibnamefont {van~den Brink}}, \bibinfo {author}
  {\bibfnamefont {G.}~\bibnamefont {Brocks}}, \ and\ \bibinfo {author}
  {\bibfnamefont {P.~J.}\ \bibnamefont {Kelly}},\ }\href {\doibase
  10.1103/PhysRevLett.99.176602} {\bibfield  {journal} {\bibinfo  {journal}
  {Phys. Rev. Lett.}\ }\textbf {\bibinfo {volume} {99}},\ \bibinfo {pages}
  {176602} (\bibinfo {year} {2007})}\BibitemShut {NoStop}%
\bibitem [{\citenamefont {Karpan}\ \emph {et~al.}(2008)\citenamefont {Karpan},
  \citenamefont {Khomyakov}, \citenamefont {Starikov}, \citenamefont
  {Giovannetti}, \citenamefont {Zwierzycki}, \citenamefont {Talanana},
  \citenamefont {Brocks}, \citenamefont {{van den Brink}},\ and\ \citenamefont
  {Kelly}}]{Karpan:prb08}%
  \BibitemOpen
  \bibfield  {author} {\bibinfo {author} {\bibfnamefont {V.~M.}\ \bibnamefont
  {Karpan}}, \bibinfo {author} {\bibfnamefont {P.~A.}\ \bibnamefont
  {Khomyakov}}, \bibinfo {author} {\bibfnamefont {A.~A.}\ \bibnamefont
  {Starikov}}, \bibinfo {author} {\bibfnamefont {G.}~\bibnamefont
  {Giovannetti}}, \bibinfo {author} {\bibfnamefont {M.}~\bibnamefont
  {Zwierzycki}}, \bibinfo {author} {\bibfnamefont {M.}~\bibnamefont
  {Talanana}}, \bibinfo {author} {\bibfnamefont {G.}~\bibnamefont {Brocks}},
  \bibinfo {author} {\bibfnamefont {J.}~\bibnamefont {{van den Brink}}}, \ and\
  \bibinfo {author} {\bibfnamefont {P.~J.}\ \bibnamefont {Kelly}},\ }\href
  {\doibase 10.1103/PhysRevB.78.195419} {\bibfield  {journal} {\bibinfo
  {journal} {Phys. Rev. B}\ }\textbf {\bibinfo {volume} {78}},\ \bibinfo
  {pages} {195419} (\bibinfo {year} {2008})}\BibitemShut {NoStop}%
\bibitem [{\citenamefont {Kryder}\ and\ \citenamefont
  {Kim}(2009)}]{Kryder:ieeem09}%
  \BibitemOpen
  \bibfield  {author} {\bibinfo {author} {\bibfnamefont {M.~H.}\ \bibnamefont
  {Kryder}}\ and\ \bibinfo {author} {\bibfnamefont {C.~S.}\ \bibnamefont
  {Kim}},\ }\href {\doibase 10.1109/TMAG.2009.2024163} {\bibfield  {journal}
  {\bibinfo  {journal} {IEEE Trans. Mag.}\ }\textbf {\bibinfo {volume} {45}},\
  \bibinfo {pages} {3406} (\bibinfo {year} {2009})}\BibitemShut {NoStop}%
\bibitem [{\citenamefont {Schmidt}\ \emph {et~al.}(2000)\citenamefont
  {Schmidt}, \citenamefont {Ferrand}, \citenamefont {Molenkamp}, \citenamefont
  {Filip},\ and\ \citenamefont {van Wees}}]{Schmidt:prb00}%
  \BibitemOpen
  \bibfield  {author} {\bibinfo {author} {\bibfnamefont {G.}~\bibnamefont
  {Schmidt}}, \bibinfo {author} {\bibfnamefont {D.}~\bibnamefont {Ferrand}},
  \bibinfo {author} {\bibfnamefont {L.~W.}\ \bibnamefont {Molenkamp}}, \bibinfo
  {author} {\bibfnamefont {A.~T.}\ \bibnamefont {Filip}}, \ and\ \bibinfo
  {author} {\bibfnamefont {B.~J.}\ \bibnamefont {van Wees}},\ }\href {\doibase
  10.1103/PhysRevB.62.R4790} {\bibfield  {journal} {\bibinfo  {journal} {Phys.
  Rev. B}\ }\textbf {\bibinfo {volume} {62}},\ \bibinfo {pages} {R4790}
  (\bibinfo {year} {2000})}\BibitemShut {NoStop}%
\bibitem [{\citenamefont {Fert}\ and\ \citenamefont
  {Jaffres}(2001)}]{Fert:prb01}%
  \BibitemOpen
  \bibfield  {author} {\bibinfo {author} {\bibfnamefont {A.}~\bibnamefont
  {Fert}}\ and\ \bibinfo {author} {\bibfnamefont {H.}~\bibnamefont {Jaffres}},\
  }\href {\doibase 10.1103/PhysRevB.64.184420} {\bibfield  {journal} {\bibinfo
  {journal} {Phys. Rev. B}\ }\textbf {\bibinfo {volume} {64}},\ \bibinfo
  {pages} {184420} (\bibinfo {year} {2001})}\BibitemShut {NoStop}%
\bibitem [{\citenamefont {Hanbicki}\ \emph {et~al.}(2002)\citenamefont
  {Hanbicki}, \citenamefont {Jonker}, \citenamefont {Itskos}, \citenamefont
  {Kioseoglou},\ and\ \citenamefont {Petrou}}]{Hanbicki:apl02}%
  \BibitemOpen
  \bibfield  {author} {\bibinfo {author} {\bibfnamefont {A.~T.}\ \bibnamefont
  {Hanbicki}}, \bibinfo {author} {\bibfnamefont {B.~T.}\ \bibnamefont
  {Jonker}}, \bibinfo {author} {\bibfnamefont {G.}~\bibnamefont {Itskos}},
  \bibinfo {author} {\bibfnamefont {G.}~\bibnamefont {Kioseoglou}}, \ and\
  \bibinfo {author} {\bibfnamefont {A.}~\bibnamefont {Petrou}},\ }\href
  {\doibase 10.1063/1.1449530} {\bibfield  {journal} {\bibinfo  {journal}
  {Appl. Phys. Lett.}\ }\textbf {\bibinfo {volume} {80}},\ \bibinfo {pages}
  {1240} (\bibinfo {year} {2002})}\BibitemShut {NoStop}%
\bibitem [{\citenamefont {Karpan}(2008)}]{Karpan08}%
  \BibitemOpen
  \bibfield  {author} {\bibinfo {author} {\bibfnamefont {V.~M.}\ \bibnamefont
  {Karpan}},\ }\emph {\bibinfo {title} {Towards perfect spin-filtering: A
  first-principles study}},\ \href {http://doc.utwente.nl/59012/} {\bibinfo
  {type} {{Ph.D.} thesis}},\ \bibinfo  {school} {University of Twente}
  (\bibinfo {year} {2008}),\ \bibinfo {note}
  {{URN:NBN:NL:UI:28-59012}}\BibitemShut {NoStop}%
\bibitem [{\citenamefont {Yazyev}\ and\ \citenamefont
  {Pasquarello}(2009)}]{Yazyev:prb09}%
  \BibitemOpen
  \bibfield  {author} {\bibinfo {author} {\bibfnamefont {O.~V.}\ \bibnamefont
  {Yazyev}}\ and\ \bibinfo {author} {\bibfnamefont {A.}~\bibnamefont
  {Pasquarello}},\ }\href {\doibase 10.1103/PhysRevB.80.035408} {\bibfield
  {journal} {\bibinfo  {journal} {Phys. Rev. B}\ }\textbf {\bibinfo {volume}
  {80}},\ \bibinfo {pages} {035408} (\bibinfo {year} {2009})}\BibitemShut
  {NoStop}%
\bibitem [{\citenamefont {Hu}\ \emph {et~al.}(2011)\citenamefont {Hu},
  \citenamefont {Yu}, \citenamefont {Zhang}, \citenamefont {Sun},\ and\
  \citenamefont {Zhong}}]{Hu:jpcc11}%
  \BibitemOpen
  \bibfield  {author} {\bibinfo {author} {\bibfnamefont {M.~L.}\ \bibnamefont
  {Hu}}, \bibinfo {author} {\bibfnamefont {Z.}~\bibnamefont {Yu}}, \bibinfo
  {author} {\bibfnamefont {K.~W.}\ \bibnamefont {Zhang}}, \bibinfo {author}
  {\bibfnamefont {L.~Z.}\ \bibnamefont {Sun}}, \ and\ \bibinfo {author}
  {\bibfnamefont {J.~X.}\ \bibnamefont {Zhong}},\ }\href {\doibase
  10.1021/jp109971r} {\bibfield  {journal} {\bibinfo  {journal} {J. Phys. Chem.
  C}\ }\textbf {\bibinfo {volume} {115}},\ \bibinfo {pages} {8260} (\bibinfo
  {year} {2011})}\BibitemShut {NoStop}%
\bibitem [{\citenamefont {Arnaud}\ \emph {et~al.}(2006)\citenamefont {Arnaud},
  \citenamefont {Lebegue}, \citenamefont {Rabiller},\ and\ \citenamefont
  {Alouani}}]{Arnaud:prl06}%
  \BibitemOpen
  \bibfield  {author} {\bibinfo {author} {\bibfnamefont {B.}~\bibnamefont
  {Arnaud}}, \bibinfo {author} {\bibfnamefont {S.}~\bibnamefont {Lebegue}},
  \bibinfo {author} {\bibfnamefont {P.}~\bibnamefont {Rabiller}}, \ and\
  \bibinfo {author} {\bibfnamefont {M.}~\bibnamefont {Alouani}},\ }\href
  {\doibase 10.1103/PhysRevLett.96.026402} {\bibfield  {journal} {\bibinfo
  {journal} {Phys. Rev. Lett.}\ }\textbf {\bibinfo {volume} {96}},\ \bibinfo
  {pages} {026402} (\bibinfo {year} {2006})}\BibitemShut {NoStop}%
\bibitem [{\citenamefont {Novoselov}\ \emph {et~al.}(2005)\citenamefont
  {Novoselov}, \citenamefont {Jiang}, \citenamefont {Schedin}, \citenamefont
  {Booth}, \citenamefont {Khotkevich}, \citenamefont {Morozov},\ and\
  \citenamefont {Geim}}]{Novoselov:pnas05}%
  \BibitemOpen
  \bibfield  {author} {\bibinfo {author} {\bibfnamefont {K.~S.}\ \bibnamefont
  {Novoselov}}, \bibinfo {author} {\bibfnamefont {D.}~\bibnamefont {Jiang}},
  \bibinfo {author} {\bibfnamefont {F.}~\bibnamefont {Schedin}}, \bibinfo
  {author} {\bibfnamefont {T.~J.}\ \bibnamefont {Booth}}, \bibinfo {author}
  {\bibfnamefont {V.~V.}\ \bibnamefont {Khotkevich}}, \bibinfo {author}
  {\bibfnamefont {S.~V.}\ \bibnamefont {Morozov}}, \ and\ \bibinfo {author}
  {\bibfnamefont {A.~K.}\ \bibnamefont {Geim}},\ }\href {\doibase
  10.1073/pnas.0502848102} {\bibfield  {journal} {\bibinfo  {journal} {Proc.
  Natl. Acad. Sci. U.S.A.}\ }\textbf {\bibinfo {volume} {102}},\ \bibinfo
  {pages} {10451} (\bibinfo {year} {2005})}\BibitemShut {NoStop}%
\bibitem [{\citenamefont {Dean}\ \emph {et~al.}(2010)\citenamefont {Dean},
  \citenamefont {Young}, \citenamefont {Meric}, \citenamefont {Lee},
  \citenamefont {Wang}, \citenamefont {Sorgenfrei}, \citenamefont {Watanabe},
  \citenamefont {Taniguchi}, \citenamefont {Kim}, \citenamefont {Shepard},\
  and\ \citenamefont {Hone}}]{Dean:natn10}%
  \BibitemOpen
  \bibfield  {author} {\bibinfo {author} {\bibfnamefont {C.~R.}\ \bibnamefont
  {Dean}}, \bibinfo {author} {\bibfnamefont {A.~F.}\ \bibnamefont {Young}},
  \bibinfo {author} {\bibfnamefont {I.}~\bibnamefont {Meric}}, \bibinfo
  {author} {\bibfnamefont {C.}~\bibnamefont {Lee}}, \bibinfo {author}
  {\bibfnamefont {L.}~\bibnamefont {Wang}}, \bibinfo {author} {\bibfnamefont
  {S.}~\bibnamefont {Sorgenfrei}}, \bibinfo {author} {\bibfnamefont
  {K.}~\bibnamefont {Watanabe}}, \bibinfo {author} {\bibfnamefont
  {T.}~\bibnamefont {Taniguchi}}, \bibinfo {author} {\bibfnamefont
  {P.}~\bibnamefont {Kim}}, \bibinfo {author} {\bibfnamefont {K.~L.}\
  \bibnamefont {Shepard}}, \ and\ \bibinfo {author} {\bibfnamefont
  {J.}~\bibnamefont {Hone}},\ }\href {\doibase 10.1038/nnano.2010.172}
  {\bibfield  {journal} {\bibinfo  {journal} {Nature Nanotechnology}\ }\textbf
  {\bibinfo {volume} {5}},\ \bibinfo {pages} {722} (\bibinfo {year}
  {2010})}\BibitemShut {NoStop}%
\bibitem [{\citenamefont {Xue}\ \emph {et~al.}(2011)\citenamefont {Xue},
  \citenamefont {Sanchez-Yamagishi}, \citenamefont {Bulmash}, \citenamefont
  {Jacquod}, \citenamefont {Deshpande}, \citenamefont {Watanabe}, \citenamefont
  {Taniguchi}, \citenamefont {Jarillo-Herrero},\ and\ \citenamefont
  {LeRoy}}]{Xue:natm11}%
  \BibitemOpen
  \bibfield  {author} {\bibinfo {author} {\bibfnamefont {J.}~\bibnamefont
  {Xue}}, \bibinfo {author} {\bibfnamefont {J.}~\bibnamefont
  {Sanchez-Yamagishi}}, \bibinfo {author} {\bibfnamefont {D.}~\bibnamefont
  {Bulmash}}, \bibinfo {author} {\bibfnamefont {P.}~\bibnamefont {Jacquod}},
  \bibinfo {author} {\bibfnamefont {A.}~\bibnamefont {Deshpande}}, \bibinfo
  {author} {\bibfnamefont {K.}~\bibnamefont {Watanabe}}, \bibinfo {author}
  {\bibfnamefont {T.}~\bibnamefont {Taniguchi}}, \bibinfo {author}
  {\bibfnamefont {P.}~\bibnamefont {Jarillo-Herrero}}, \ and\ \bibinfo {author}
  {\bibfnamefont {B.~J.}\ \bibnamefont {LeRoy}},\ }\href {\doibase
  10.1038/nmat2968} {\bibfield  {journal} {\bibinfo  {journal} {Nature
  Materials}\ }\textbf {\bibinfo {volume} {10}},\ \bibinfo {pages} {282}
  (\bibinfo {year} {2011})}\BibitemShut {NoStop}%
\bibitem [{\citenamefont {Decker}\ \emph {et~al.}(2011)\citenamefont {Decker},
  \citenamefont {Wang}, \citenamefont {Brar}, \citenamefont {Regan},
  \citenamefont {Tsai}, \citenamefont {Wu}, \citenamefont {Gannett},
  \citenamefont {Zettl},\ and\ \citenamefont {Crommie}}]{Decker:nanol11}%
  \BibitemOpen
  \bibfield  {author} {\bibinfo {author} {\bibfnamefont {R.}~\bibnamefont
  {Decker}}, \bibinfo {author} {\bibfnamefont {Y.}~\bibnamefont {Wang}},
  \bibinfo {author} {\bibfnamefont {V.~W.}\ \bibnamefont {Brar}}, \bibinfo
  {author} {\bibfnamefont {W.}~\bibnamefont {Regan}}, \bibinfo {author}
  {\bibfnamefont {H.-Z.}\ \bibnamefont {Tsai}}, \bibinfo {author}
  {\bibfnamefont {Q.}~\bibnamefont {Wu}}, \bibinfo {author} {\bibfnamefont
  {W.}~\bibnamefont {Gannett}}, \bibinfo {author} {\bibfnamefont
  {A.}~\bibnamefont {Zettl}}, \ and\ \bibinfo {author} {\bibfnamefont {M.~F.}\
  \bibnamefont {Crommie}},\ }\href {\doibase 10.1021/nl2005115} {\bibfield
  {journal} {\bibinfo  {journal} {Nano Letters}\ }\textbf {\bibinfo {volume}
  {11}},\ \bibinfo {pages} {2291} (\bibinfo {year} {2011})}\BibitemShut
  {NoStop}%
\bibitem [{\citenamefont {Giovannetti}\ \emph {et~al.}(2007)\citenamefont
  {Giovannetti}, \citenamefont {Khomyakov}, \citenamefont {Brocks},
  \citenamefont {Kelly},\ and\ \citenamefont {van~den
  Brink}}]{Giovannetti:prb07}%
  \BibitemOpen
  \bibfield  {author} {\bibinfo {author} {\bibfnamefont {G.}~\bibnamefont
  {Giovannetti}}, \bibinfo {author} {\bibfnamefont {P.~A.}\ \bibnamefont
  {Khomyakov}}, \bibinfo {author} {\bibfnamefont {G.}~\bibnamefont {Brocks}},
  \bibinfo {author} {\bibfnamefont {P.~J.}\ \bibnamefont {Kelly}}, \ and\
  \bibinfo {author} {\bibfnamefont {J.}~\bibnamefont {van~den Brink}},\ }\href
  {\doibase 10.1103/PhysRevB.76.073103} {\bibfield  {journal} {\bibinfo
  {journal} {Phys. Rev. B}\ }\textbf {\bibinfo {volume} {76}},\ \bibinfo
  {pages} {073103} (\bibinfo {year} {2007})}\BibitemShut {NoStop}%
\bibitem [{\citenamefont {Bl{\"{o}}chl}(1994)}]{Blochl:prb94b}%
  \BibitemOpen
  \bibfield  {author} {\bibinfo {author} {\bibfnamefont {P.~E.}\ \bibnamefont
  {Bl{\"{o}}chl}},\ }\href {\doibase 10.1103/PhysRevB.49.16223} {\bibfield
  {journal} {\bibinfo  {journal} {Phys. Rev. B}\ }\textbf {\bibinfo {volume}
  {50}},\ \bibinfo {pages} {17953} (\bibinfo {year} {1994})}\BibitemShut
  {NoStop}%
\bibitem [{\citenamefont {Kresse}\ and\ \citenamefont
  {Hafner}(1993)}]{Kresse:prb93}%
  \BibitemOpen
  \bibfield  {author} {\bibinfo {author} {\bibfnamefont {G.}~\bibnamefont
  {Kresse}}\ and\ \bibinfo {author} {\bibfnamefont {J.}~\bibnamefont
  {Hafner}},\ }\href {\doibase 10.1103/PhysRevB.47.558} {\bibfield  {journal}
  {\bibinfo  {journal} {Phys. Rev. B}\ }\textbf {\bibinfo {volume} {47}},\
  \bibinfo {pages} {558} (\bibinfo {year} {1993})}\BibitemShut {NoStop}%
\bibitem [{\citenamefont {Kresse}\ and\ \citenamefont
  {Furthm{\"{u}}ller}(1996)}]{Kresse:prb96}%
  \BibitemOpen
  \bibfield  {author} {\bibinfo {author} {\bibfnamefont {G.}~\bibnamefont
  {Kresse}}\ and\ \bibinfo {author} {\bibfnamefont {J.}~\bibnamefont
  {Furthm{\"{u}}ller}},\ }\href {\doibase 10.1103/PhysRevB.54.11169} {\bibfield
   {journal} {\bibinfo  {journal} {Phys. Rev. B}\ }\textbf {\bibinfo {volume}
  {54}},\ \bibinfo {pages} {11169} (\bibinfo {year} {1996})}\BibitemShut
  {NoStop}%
\bibitem [{\citenamefont {Kresse}\ and\ \citenamefont
  {Joubert}(1999)}]{Kresse:prb99}%
  \BibitemOpen
  \bibfield  {author} {\bibinfo {author} {\bibfnamefont {G.}~\bibnamefont
  {Kresse}}\ and\ \bibinfo {author} {\bibfnamefont {D.}~\bibnamefont
  {Joubert}},\ }\href {\doibase 10.1103/PhysRevB.59.1758} {\bibfield  {journal}
  {\bibinfo  {journal} {Phys. Rev. B}\ }\textbf {\bibinfo {volume} {59}},\
  \bibinfo {pages} {1758} (\bibinfo {year} {1999})}\BibitemShut {NoStop}%
\bibitem [{\citenamefont {Xia}\ \emph {et~al.}(2006)\citenamefont {Xia},
  \citenamefont {Zwierzycki}, \citenamefont {Talanana}, \citenamefont {Kelly},\
  and\ \citenamefont {Bauer}}]{Xia:prb06}%
  \BibitemOpen
  \bibfield  {author} {\bibinfo {author} {\bibfnamefont {K.}~\bibnamefont
  {Xia}}, \bibinfo {author} {\bibfnamefont {M.}~\bibnamefont {Zwierzycki}},
  \bibinfo {author} {\bibfnamefont {M.}~\bibnamefont {Talanana}}, \bibinfo
  {author} {\bibfnamefont {P.~J.}\ \bibnamefont {Kelly}}, \ and\ \bibinfo
  {author} {\bibfnamefont {G.~E.~W.}\ \bibnamefont {Bauer}},\ }\href {\doibase
  10.1103/PhysRevB.73.064420} {\bibfield  {journal} {\bibinfo  {journal} {Phys.
  Rev. B}\ }\textbf {\bibinfo {volume} {73}},\ \bibinfo {pages} {064420}
  (\bibinfo {year} {2006})}\BibitemShut {NoStop}%
\bibitem [{\citenamefont {Butler}\ \emph {et~al.}(2001)\citenamefont {Butler},
  \citenamefont {Zhang}, \citenamefont {Schulthess},\ and\ \citenamefont
  {MacLaren}}]{Butler:prb01a}%
  \BibitemOpen
  \bibfield  {author} {\bibinfo {author} {\bibfnamefont {W.~H.}\ \bibnamefont
  {Butler}}, \bibinfo {author} {\bibfnamefont {X.-G.}\ \bibnamefont {Zhang}},
  \bibinfo {author} {\bibfnamefont {T.~C.}\ \bibnamefont {Schulthess}}, \ and\
  \bibinfo {author} {\bibfnamefont {J.~M.}\ \bibnamefont {MacLaren}},\ }\href
  {\doibase 10.1103/PhysRevB.63.054416} {\bibfield  {journal} {\bibinfo
  {journal} {Phys. Rev. B}\ }\textbf {\bibinfo {volume} {63}},\ \bibinfo
  {pages} {054416} (\bibinfo {year} {2001})}\BibitemShut {NoStop}%
\bibitem [{\citenamefont {Mathon}\ and\ \citenamefont
  {Umerski}(2001)}]{Mathon:prb01}%
  \BibitemOpen
  \bibfield  {author} {\bibinfo {author} {\bibfnamefont {J.}~\bibnamefont
  {Mathon}}\ and\ \bibinfo {author} {\bibfnamefont {A.}~\bibnamefont
  {Umerski}},\ }\href {\doibase 10.1103/PhysRevB.63.220403} {\bibfield
  {journal} {\bibinfo  {journal} {Phys. Rev. B}\ }\textbf {\bibinfo {volume}
  {63}},\ \bibinfo {pages} {220403(R)} (\bibinfo {year} {2001})}\BibitemShut
  {NoStop}%
\bibitem [{\citenamefont {Parkin}\ \emph {et~al.}(2004)\citenamefont {Parkin},
  \citenamefont {Kaiser}, \citenamefont {Panchula}, \citenamefont {Rice},
  \citenamefont {Hughes}, \citenamefont {Samant},\ and\ \citenamefont
  {Yang}}]{Parkin:natm04}%
  \BibitemOpen
  \bibfield  {author} {\bibinfo {author} {\bibfnamefont {S.~S.~P.}\
  \bibnamefont {Parkin}}, \bibinfo {author} {\bibfnamefont {C.}~\bibnamefont
  {Kaiser}}, \bibinfo {author} {\bibfnamefont {A.}~\bibnamefont {Panchula}},
  \bibinfo {author} {\bibfnamefont {P.~M.}\ \bibnamefont {Rice}}, \bibinfo
  {author} {\bibfnamefont {B.}~\bibnamefont {Hughes}}, \bibinfo {author}
  {\bibfnamefont {M.}~\bibnamefont {Samant}}, \ and\ \bibinfo {author}
  {\bibfnamefont {S.~H.}\ \bibnamefont {Yang}},\ }\href {\doibase
  10.1038/nmat1256} {\bibfield  {journal} {\bibinfo  {journal} {Nature
  Materials}\ }\textbf {\bibinfo {volume} {3}},\ \bibinfo {pages} {862}
  (\bibinfo {year} {2004})}\BibitemShut {NoStop}%
\bibitem [{\citenamefont {Yuasa}\ \emph {et~al.}(2004)\citenamefont {Yuasa},
  \citenamefont {Nagahama}, \citenamefont {Fukushima}, \citenamefont {Suzuki},\
  and\ \citenamefont {Ando}}]{Yuasa:natm04}%
  \BibitemOpen
  \bibfield  {author} {\bibinfo {author} {\bibfnamefont {S.}~\bibnamefont
  {Yuasa}}, \bibinfo {author} {\bibfnamefont {T.}~\bibnamefont {Nagahama}},
  \bibinfo {author} {\bibfnamefont {A.}~\bibnamefont {Fukushima}}, \bibinfo
  {author} {\bibfnamefont {Y.}~\bibnamefont {Suzuki}}, \ and\ \bibinfo {author}
  {\bibfnamefont {K.}~\bibnamefont {Ando}},\ }\href {\doibase 10.1038/nmat1257}
  {\bibfield  {journal} {\bibinfo  {journal} {Nature Materials}\ }\textbf
  {\bibinfo {volume} {3}},\ \bibinfo {pages} {868} (\bibinfo {year}
  {2004})}\BibitemShut {NoStop}%
\bibitem [{\citenamefont {Jedema}\ \emph {et~al.}(2002)\citenamefont {Jedema},
  \citenamefont {Heersche}, \citenamefont {Filip}, \citenamefont {Baselmans},\
  and\ \citenamefont {van Wees}}]{Jedema:nat02}%
  \BibitemOpen
  \bibfield  {author} {\bibinfo {author} {\bibfnamefont {F.~J.}\ \bibnamefont
  {Jedema}}, \bibinfo {author} {\bibfnamefont {H.~B.}\ \bibnamefont
  {Heersche}}, \bibinfo {author} {\bibfnamefont {A.~T.}\ \bibnamefont {Filip}},
  \bibinfo {author} {\bibfnamefont {J.~J.~A.}\ \bibnamefont {Baselmans}}, \
  and\ \bibinfo {author} {\bibfnamefont {B.~J.}\ \bibnamefont {van Wees}},\
  }\href {\doibase 10.1038/416713a} {\bibfield  {journal} {\bibinfo  {journal}
  {Nature}\ }\textbf {\bibinfo {volume} {406}},\ \bibinfo {pages} {713}
  (\bibinfo {year} {2002})}\BibitemShut {NoStop}%
\bibitem [{\citenamefont {Giovannetti}\ \emph {et~al.}(2008)\citenamefont
  {Giovannetti}, \citenamefont {Khomyakov}, \citenamefont {Brocks},
  \citenamefont {Karpan}, \citenamefont {van~den Brink},\ and\ \citenamefont
  {Kelly}}]{Giovannetti:prl08}%
  \BibitemOpen
  \bibfield  {author} {\bibinfo {author} {\bibfnamefont {G.}~\bibnamefont
  {Giovannetti}}, \bibinfo {author} {\bibfnamefont {P.~A.}\ \bibnamefont
  {Khomyakov}}, \bibinfo {author} {\bibfnamefont {G.}~\bibnamefont {Brocks}},
  \bibinfo {author} {\bibfnamefont {V.~M.}\ \bibnamefont {Karpan}}, \bibinfo
  {author} {\bibfnamefont {J.}~\bibnamefont {van~den Brink}}, \ and\ \bibinfo
  {author} {\bibfnamefont {P.~J.}\ \bibnamefont {Kelly}},\ }\href {\doibase
  10.1103/PhysRevLett.101.026803} {\bibfield  {journal} {\bibinfo  {journal}
  {Phys. Rev. Lett.}\ }\textbf {\bibinfo {volume} {101}},\ \bibinfo {pages}
  {026803} (\bibinfo {year} {2008})}\BibitemShut {NoStop}%
\bibitem [{\citenamefont {Khomyakov}\ \emph {et~al.}(2009)\citenamefont
  {Khomyakov}, \citenamefont {Giovannetti}, \citenamefont {Rusu}, \citenamefont
  {Brocks}, \citenamefont {van~den Brink},\ and\ \citenamefont
  {Kelly}}]{Khomyakov:prb09}%
  \BibitemOpen
  \bibfield  {author} {\bibinfo {author} {\bibfnamefont {P.~A.}\ \bibnamefont
  {Khomyakov}}, \bibinfo {author} {\bibfnamefont {G.}~\bibnamefont
  {Giovannetti}}, \bibinfo {author} {\bibfnamefont {P.~C.}\ \bibnamefont
  {Rusu}}, \bibinfo {author} {\bibfnamefont {G.}~\bibnamefont {Brocks}},
  \bibinfo {author} {\bibfnamefont {J.}~\bibnamefont {van~den Brink}}, \ and\
  \bibinfo {author} {\bibfnamefont {P.~J.}\ \bibnamefont {Kelly}},\ }\href
  {\doibase 10.1103/PhysRevB.79.195425} {\bibfield  {journal} {\bibinfo
  {journal} {Phys. Rev. B}\ }\textbf {\bibinfo {volume} {79}},\ \bibinfo
  {pages} {195425} (\bibinfo {year} {2009})}\BibitemShut {NoStop}%
\bibitem [{\citenamefont {Watanabe}\ \emph {et~al.}(1996)\citenamefont
  {Watanabe}, \citenamefont {Itoh}, \citenamefont {Sasaki},\ and\ \citenamefont
  {Mizushima}}]{Watanabe:prl96}%
  \BibitemOpen
  \bibfield  {author} {\bibinfo {author} {\bibfnamefont {M.~O.}\ \bibnamefont
  {Watanabe}}, \bibinfo {author} {\bibfnamefont {S.}~\bibnamefont {Itoh}},
  \bibinfo {author} {\bibfnamefont {T.}~\bibnamefont {Sasaki}}, \ and\ \bibinfo
  {author} {\bibfnamefont {K.}~\bibnamefont {Mizushima}},\ }\href {\doibase
  10.1103/PhysRevLett.77.187} {\bibfield  {journal} {\bibinfo  {journal} {Phys.
  Rev. Lett.}\ }\textbf {\bibinfo {volume} {77}},\ \bibinfo {pages} {187}
  (\bibinfo {year} {1996})}\BibitemShut {NoStop}%
\bibitem [{\citenamefont {Chen}\ \emph {et~al.}(1999)\citenamefont {Chen},
  \citenamefont {Barnard}, \citenamefont {Palmer}, \citenamefont {Watanabe},\
  and\ \citenamefont {Sasaki}}]{Chen:prl99}%
  \BibitemOpen
  \bibfield  {author} {\bibinfo {author} {\bibfnamefont {Y.}~\bibnamefont
  {Chen}}, \bibinfo {author} {\bibfnamefont {J.~C.}\ \bibnamefont {Barnard}},
  \bibinfo {author} {\bibfnamefont {R.~E.}\ \bibnamefont {Palmer}}, \bibinfo
  {author} {\bibfnamefont {M.~O.}\ \bibnamefont {Watanabe}}, \ and\ \bibinfo
  {author} {\bibfnamefont {T.}~\bibnamefont {Sasaki}},\ }\href {\doibase
  10.1103/PhysRevLett.83.2406} {\bibfield  {journal} {\bibinfo  {journal}
  {Phys. Rev. Lett.}\ }\textbf {\bibinfo {volume} {83}},\ \bibinfo {pages}
  {2406} (\bibinfo {year} {1999})}\BibitemShut {NoStop}%
\bibitem [{\citenamefont {Zwierzycki}\ \emph {et~al.}(2003)\citenamefont
  {Zwierzycki}, \citenamefont {Xia}, \citenamefont {Kelly}, \citenamefont
  {Bauer},\ and\ \citenamefont {Turek}}]{Zwierzycki:prb03}%
  \BibitemOpen
  \bibfield  {author} {\bibinfo {author} {\bibfnamefont {M.}~\bibnamefont
  {Zwierzycki}}, \bibinfo {author} {\bibfnamefont {K.}~\bibnamefont {Xia}},
  \bibinfo {author} {\bibfnamefont {P.~J.}\ \bibnamefont {Kelly}}, \bibinfo
  {author} {\bibfnamefont {G.~E.~W.}\ \bibnamefont {Bauer}}, \ and\ \bibinfo
  {author} {\bibfnamefont {I.}~\bibnamefont {Turek}},\ }\href {\doibase
  10.1103/PhysRevB.67.092401} {\bibfield  {journal} {\bibinfo  {journal} {Phys.
  Rev. B}\ }\textbf {\bibinfo {volume} {67}},\ \bibinfo {pages} {092401}
  (\bibinfo {year} {2003})}\BibitemShut {NoStop}%
\bibitem [{\citenamefont {Xu}\ \emph {et~al.}(2006)\citenamefont {Xu},
  \citenamefont {Karpan}, \citenamefont {Xia}, \citenamefont {Zwierzycki},
  \citenamefont {Marushchenko},\ and\ \citenamefont {Kelly}}]{Xu:prb06a}%
  \BibitemOpen
  \bibfield  {author} {\bibinfo {author} {\bibfnamefont {P.~X.}\ \bibnamefont
  {Xu}}, \bibinfo {author} {\bibfnamefont {V.~M.}\ \bibnamefont {Karpan}},
  \bibinfo {author} {\bibfnamefont {K.}~\bibnamefont {Xia}}, \bibinfo {author}
  {\bibfnamefont {M.}~\bibnamefont {Zwierzycki}}, \bibinfo {author}
  {\bibfnamefont {I.}~\bibnamefont {Marushchenko}}, \ and\ \bibinfo {author}
  {\bibfnamefont {P.~J.}\ \bibnamefont {Kelly}},\ }\href {\doibase
  10.1103/PhysRevB.73.180402} {\bibfield  {journal} {\bibinfo  {journal} {Phys.
  Rev. B}\ }\textbf {\bibinfo {volume} {73}},\ \bibinfo {pages} {180402(R)}
  (\bibinfo {year} {2006})}\BibitemShut {NoStop}%
\end{thebibliography}

\end{document}